# Simulation of optoelectronic oscillator injection locking, pulling & spiking phenomena

Abhijit Banerjee[1*], Trevor James Hall[2], *Senior Member, IEEE*

[1]Academy of Technology, Adisaptagram, Hooghly, West Bengal, India

[2]Photonics Technology Laboratory, University of Ottawa, Advanced Research Complex, 25 Templeton Street, Ottawa, Ontario, K1N 6N5, Canada.

*abhijit.banerjee@aot.edu.in

ABSTRACT

Complex envelope and reduced phase simulation models describing the dynamical behaviour of an optoelectronic oscillator (OEO) under injection by an external source are described. The models build on the foundations of a previously reported delay integral / differential equation (DDE) theory of injection locking of time delay oscillators (TDO) such as the OEO. The DDE formulation is particularly amenable to high precision simulation using the Simulink™ block diagram environment. The correspondence between the blocks and the oscillator components offers intuition and considerable freedom to explore different circuit architectures and design variations with minimal coding effort. The simulations facilitate the study of the profound effect the multimode nature of a TDO has on its dynamical behavior. The reduced phase models that make use of the Leeson approximation are generally successful in reproducing the results of complex envelope models for established oscillations except for spiking phenomena for which the Leeson approximation fails. Simulation results demonstrating phenomena not captured by classical injection theory are presented, including multimode oscillation, the appearance of sidemodes in the RF and phase noise spectrum, and persistent spike trains redolent of recent experimental observations of $2\pi$ phase pulse trains in a broadband OEO under injection.

I. INTRODUCTION

Clock jitter or, equivalently, oscillator phase / frequency fluctuation limits the precision of time and frequency measurements, which is of major consequence in applications such as optical & wireless communications, high speed digital electronics, radar & lidar, astronomy and terahertz technology. RF photonic technology offers the potential to realise oscillators with phase noise levels orders of magnitude lower than conventional sources at frequencies from the microwave to the terahertz region of the spectrum. Among a variety of means using photonic technology to generate pristine RF carriers, the optoelectronic oscillator (OEO) is the most suited to practical deployment. Reference [1] provides a recent review of the extensive literature that has arisen following the introduction of the OEO [2].

A laser and OEO are examples of a time delay oscillator (TDO). The laser generates optical carriers using a resonant cavity containing the sustaining amplifier. The OEO generates microwave carriers using an RF photonic link consisting of laser; optical intensity modulator; optical fibre; photo-receiver; RF amplifier and filter, which drives the modulator, closing the loop and sustaining oscillation (see Fig. 1). The low loss of optical fibre (~0.2 dB/km) permits delay line lengths of ~ 5 km offering, for example, OEO phase noise performance of -145 dBc/Hz at 10 kHz offset from a 10 GHz carrier [3]. However, the frequency interval between adjacent oscillation modes decreases in inverse proportion to the delay (40 kHz for 5 km), and a practical RF resonator cannot provide sufficient selectivity to suppress the multitude of sidemode resonances. Multimode operation is an artefact of an OEO with profound consequences on its behaviour [3,4].

There is a conceptual distinction between a single-mode or multimode oscillation and a single-mode or multimode oscillator. In applications as a low noise oscillator, evolution to a single-mode oscillation state is a desirable property but it does not change the nature of the oscillator from a multimode to a single-mode oscillator. Modes correspond to the attractive states (limit cycles) of the equations of motion that govern the oscillator. Whether the evolution of the oscillation reaches an attractive state, and, if so, which of the multitude of attractive states becomes the final state of oscillation depends upon both the equations of motion and the initial condition. Even when the initial condition corresponds to a single-mode oscillation state the presence of multiple attractors is manifest by any perturbation of a multimode oscillator including: fluctuations & noise, which excite sidemodes and give rise to spurious resonances in the phase noise spectrum; intra-loop phase modulation, especially at frequencies resonant with the intermodal frequency that gives rise to giant phase modulation, frequency combs & mode-locking; tuning scans and transients that, if rapid, excite sidemodes or, if slow, may cause mode-hops. In addition, giant phase modulation is a source of a modulational instability in a phase lock loop (PLL) controlled OEO [3].

A new delay integral/differential equation (DDE) formulation of an OEO under external injection is presented in a prior work [4] that removes the implicit assumption of a single-mode oscillator under weak injection made in previous treatments that apply classical injection locking theory to OEOs. The emphasis of the prior work is on the theory and the correct prediction of the experimentally observed phase noise spectrum; simulation results were precluded by space considerations. This work corrects that omission. The DDE formulation is particularly amenable to high precision simulation using the Simulink™ block diagram environment. The correspondence between blocks and the component of the oscillator offers intuition into the behaviour of the oscillator and considerable freedom to explore different circuit architectures and design variations with minimal coding effort. The simulations are valuable in illustrating the dependence on system parameters of behaviour that can be rich with phenomena that defy analytic solution.

Levy et al [5] presents a ground-breaking numerical simulation model based on a multiple timescale approach to study mode competition during the OEO initial start-up regime as well as the amplitude and phase of the established oscillation. This model is further extended in [6] to accurately determine the phase noise in single loop and dual-loop OEOs and its dependence on parameters. Mikitchuk et al [7] reports a numerical nonlinear time varying model of a delay line OEO, which can simulate the stationary behavior and dynamical instabilities in single-loop OEO and multiloop OEO. More recently, Yuan et. al. [8] reports a time domain convolution simulation model to study the real time dynamic process in the initial and established oscillation regimes of an OEO under external injection. The convolution describes the time domain response in terms of the impulse response of the RF bandpass filter (BPF) and the method of steps for solving a DDE is used to calculate the current segment of data over the delay interval from the previous segment with appropriate account for the convolution extending across segments.

In these works, the nonlinear transfer function of a Mach-Zehnder modulator (MZM) is assumed to be responsible for the gain control mechanism. This introduces a first order Bessel function dependence of the saturated gain, the non-monotonicity of which is responsible for an MZM overdrive envelope instability [5,9]. Complex dynamical instabilities are of theoretical interest and have been thoroughly investigate by Chembo et al [9–12]. However, in low-noise oscillator applications, it is preferable that the MZM is driven between adjacent minimum and maximum transmission points and no further to avoid this instability. That is, RF amplifier limiting should be the principal source of gain saturation and modelled appropriately (see Section II.B(i)).

OEO simulation tools are generally implemented using custom code. Belkin et al [13] is a first attempt to use a commercial photonic circuit simulation tool (VPI Transmission Maker™) which is advantageous in offering a comprehensive library of optical and some electronic component models. However, the simulation must adequately sample the RF modulation of the optical carrier which severely restricts the fibre delay line length that can be modelled. The Simulink™ models described herein treat the RF-photonic link as an RF component (a transport delay) so adequate sampling is required only of the fluctuations of the complex envelope reducing the number of samples held within the delay line by several orders of magnitude.

The paper describes for the first time an implementation of a comprehensive simulation model of an OEO under injection by an external source using the Simulink™ block diagram environment. Simulation results reveal phenomena not captured by classical single-mode injection locking theory. Specifically, spiking phenomena are observed that are related to a serrodyne mode injection-pulling solution of an idealised reduced phase-only model. The formation of a periodic spike train necessarily requires the *coherent* superposition of a multitude of modes and, as such, spiking is a manifestation of mode-locking. TDOs such as lasers and OEOs support a multitude of modes and consequently, spiking is to be expected. The paper reports the first observation in simulation of mode-locking by injection induced intra-loop phase modulation and is redolent of recent experimental observations reported by Diakonov et al [14]. of $2\pi$ phase pulse trains generated by a broadband OEO under injection Spiking is of interest for the emulation of excitability in neuroscience [15] and its application to high-speed neuromorphic computing [16]. Mode-locked laser theory and practice is well advanced [17] and analogous methods of mode-locking an OEO are being explored for application to pulsed radar systems [18-22].

The structure of the paper is as follows. In Section II a brief review is presented of the analytical model of a free OEO and an injection driven OEO in the complex envelope representation, including an explanation of how amplifier gain saturation by limiting is modelled. In addition, the reduction to a phase-only model of an OEO under injection and its phase-locked solutions are described. A special 'serrodyne' quasi-locked state unique to a broadband TDO is introduced for the first time. Section III describes the detailed implementation of the envelope and phase models in the Simulink™ block diagram environment. Section IV then presents representative results generated by the models that illustrate the multimode nature of the oscillator and its profound effect on its time domain behaviour and spectral features. Section V concludes the paper with a summary of the principal findings.

## II. ANALYTICAL MODEL

The OEO under study, depicted in Fig. 1, consists of an RF photonic link and an RF amplifier chain configured into a loop. The laser followed by the MZM converts the RF signal into an intensity-modulated optical carrier. A single-mode optical fibre coil with delay $\tau_D$ is used as the optical fibre delay line (OFDL). The RF modulation is recovered by the photodetector (PD). The combination of a laser, an MZM, an OFDL and a PD is functionally equivalent to an RF-photonic link, which is used to provide the time delay of the oscillator. The sustaining amplifier of the oscillator is provided by the RF chain. The RF chain consists of an RF amplifier followed by an RF BPF, also referred to as an RF resonator, which has an on-resonance group delay $\tau_R$. The round-trip time is $\tau_G = \tau_D + \tau_R$ and normally $\tau_D \gg \tau_R$.

The OEO achieves low phase noise through the large delay enabled by the exceptionally low transmission loss of optical fibre. However, the frequency interval between potential oscillation modes is inversely proportional to the delay. Typically, an RF BPF is concatenated with the delay line to select the desired oscillation mode. At microwave frequencies the passband of the RF filter is broad relative to the frequency interval between potential oscillation modes; and spurious resonances due to sidemodes falling within the RF filter passband are clearly seen both in the RF and phase noise spectral densities.

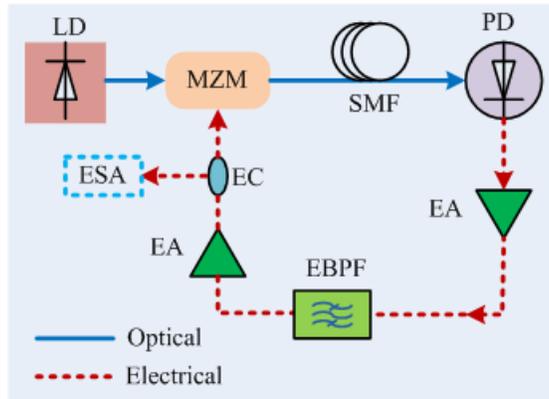

Figure 1. Schematic of a single-loop OEO. LD: semiconductor laser diode; MZM: Mach-Zehnder Modulator; SMF: single mode fibre, OFDL: optical fibre delay line; PD: photo detector, EA: electronic amplifier (RF amplifier); EBPF: electronic bandpass filter (BPF); EC: electronic coupler; ESA: electronic spectrum analyser.

The RF BPF plays an important role in promoting single-mode oscillation. Initially the oscillation builds up from an initial transient or noise and the sustaining amplifier operates in its linear regime. A potential oscillation mode grows at a rate proportional to the net gain it experiences on each round-trip. A suitably tuned RF BPF with a bell-shaped transmission function magnitude favours the growth of one mode slightly above all others. The gain control mechanism, usually saturation of the sustaining amplifier, reduces the round-trip gain as the magnitude of the oscillation grows until the favoured mode is sustained by net unit gain and neighbouring modes experience net loss and slowly decay.

Once the oscillation is established, the gain control mechanism holds the magnitude of the oscillation substantially constant, but the phase of the oscillation continues to be modulated by the residual sidemode spectral components. The RF BPF then acts as a phase filter suppressing the phase modulation and thereby continuing the decay of the sidemode spectral components. In this regime, the step response of the oscillation phase has a staircase waveform. The RF filter smooths the top step of the phase staircase each round-trip and the staircase evolves ultimately into a linear ramp, i.e., a step change in intra-loop phase ultimately results in a change in frequency of a pristine oscillation. Perturbations due to a variety of fluctuations within the oscillator maintain a residual level of sidemode spectral components that otherwise would decay asymptotically to zero.

*A. Free running single-loop OEO*

The dynamical equation of a free running single loop OEO is given by:

$$v = e^{i\phi} K\left(h \otimes \left(D_{\tau_D} v\right)\right) \tag{1}$$

where $v$ is the complex envelope of the oscillation at the output of the RF amplifier following the RF BPF and $\phi$ represents the net phase contributed by an intra-loop tuning element and other components. The action of the RF BPF on the complex envelope is represented by a convolution by the impulse response $h$ of the equivalent baseband low pass filter (LPF) [23]. The sustaining amplifier is represented by the operator $K$, which is characterized by a real valued large signal gain $\kappa$ that decreases from the linear gain in such away that the magnitude of the complex envelope of the amplifier output is held substantially constant when the amplifier operates in saturation: A more precise characterisation is given in Section II.B(i). The operator $D_{\tau_D}$ represents a delay:

$$\left(D_{\tau_D} v\right)(t) = v(t - \tau_D) \tag{2}$$

Eq. (1) admits a family of freely oscillating mode solutions:

$$v(t) = a_p \exp(s_p t) \quad ; \quad s_p = \sigma_p + i\omega_p \tag{3}$$

subject to the complex Barkhausen condition:

$$\kappa H(s_p) e^{i\phi} \exp(-s_p \tau_D) = 1 \implies \begin{cases} \kappa |H(s_p)| \exp(-\sigma_p \tau_D) = 1 \\ \omega_p \tau_D = 2p\pi + \phi + \arg\left(H(s_p)\right) \quad p \in \mathbb{Z} \end{cases} \tag{4}$$

The magnitude of a single freely oscillating mode close to the passband centre frequency of the RF BPF remains constant as time progresses if $\kappa = 1$, which will be maintained by the gain control mechanism. The integer $p$ represents the additional number of cycles contained within the delay line of mode $p$ relative to the reference mode $p = 0$. The frequency interval between allowed freely oscillating modes defines the free-spectral range (FSR), i.e., the change in frequency for a unit increment of the number of cycles within the loop. The oscillator is tuned over an FSR for $\phi \in (-\pi, \pi]$.

*B. Single-loop OEO under RF signal injection*

### (i) Complex envelope model

The addition of a complex envelope $w$ to the right-hand side of (1) introduces a forcing term representing the oscillator under injection. The resulting complex envelope model of injection locking of a time delay oscillator is conveniently written in a form:

$$\begin{aligned} u &= v + w \\ v &= e^{i\phi} K\left(h \otimes \left(D_{\tau_D} u\right)\right) \end{aligned} \tag{5}$$

that highlights the role of the linear superposition $u$ of the complex envelopes representing the oscillation $v$ and the injected carrier $w$.

To describe gain saturation by limiting of an RF amplifier driving an MZM, it is assumed that each nonlinear stage is followed by a BPF that dissipates of all intermodulation products but has a sufficiently broad passband relative to the signal bandwidth that the phase of the signal is passed undisturbed. Consequently, with appropriate choice of time co-ordinate, the signal $x$ may be represented locally (on the scale of a period), as a pure carrier:

$$x(t) = a \cos(t) \tag{6}$$

where $a$ is a locally constant magnitude. Since $x$ is even and periodic one can restrict attention to the interval $t \in [0, \pi/2]$. Fast limiting at unit amplitude may be described by:

$$\bar{x}(t) = x(t) + y(t) \tag{7}$$

where $y(t)$ accounts for all distortion:

$$y(t) = \begin{cases} 0 & ; \quad t \in [\xi, \pi/2] \\ 1 - x(t) & ; \quad t \in (0, \xi) \end{cases} \tag{8}$$

and $\xi$ defines the interval over which limiting occurs.

$$\begin{aligned} \xi = 0 & \quad ; \quad a \leq 1 \\ a \cos(\xi) = 1 & \quad ; \quad a > 1 \end{aligned} \tag{9}$$

The amplitude $\tilde{a}$ of the fundamental harmonic of $\bar{x}$ is:

$$\tilde{a} = a + \frac{4}{\pi} \int_0^\xi y(t) \cos(t) \, dt = a \left( 1 - \frac{2}{\pi} \int_0^\xi dt - \frac{2}{\pi} \int_0^\xi \cos(2t) \, dt \right) + \frac{4}{\pi} \int_0^\xi \cos(t) \, dt \tag{10}$$

Evaluating the integrals leads to:

$$\tilde{a} = \begin{cases} a & ; \quad a \leq 1 \\ a(1 - (2\xi - \sin(2\xi))/\pi) & ; \quad a > 1 \end{cases} \tag{11}$$

$\tilde{a}$ is a continuous monotonic function of $a$ which in the hard-clipping limit tends to $4/\pi$ corresponding to the amplitude of the fundamental harmonic of a unit amplitude square wave. The saturation limit may be normalised to unity while retaining a unit unsaturated gain by pre-multiplication by $\pi/4$ and post-multiplication by $4/\pi$. The saturated gain $\kappa = \tilde{a}/a$ is the describing function [24] of the describing function method, a special case of the harmonic balance method [25]; in which a truncated Fourier series representation is used to approximate periodic solutions of a nonlinear system. The harmonic balance method is a special case of the mathematically rigorous Galerkin method [26].

### (ii) Reduced phase-only model

The magnitude of an established oscillation is substantially maintained constant by the saturation of the sustaining amplifier. Amplitude fluctuations are highly suppressed. It is consequently an excellent approximation in the established oscillation regime to reduce the complex envelope model to a reduced phase-only model given by:

$$\theta_v = \phi + h \otimes \left( D_{\tau_D} \left( \tan^{-1} \left( \frac{\rho \sin(\theta_w - \theta_v)}{1 + \rho \cos(\theta_w - \theta_v)} \right) + \theta_v \right) \right) \tag{12}$$

where $\theta_v$ is the oscillator phase; $\theta_w$ is the phase of the injected carrier; and $\rho$ is the ratio of the magnitudes of the injected carrier and oscillation. This model invokes the Leeson approximation [27,28] of the action of a BPF on the phase of a complex envelope.

The arctangent function is to be understood in the sense:

$$\theta = \tan^{-1}(y/x) = \arg(z) \quad ; \quad z = x + iy \tag{13}$$

It is a multivalued function with an infinity of branches separated by $2\pi$. Each branch corresponds to one of an infinity of oscillation modes. For $\rho < 1$ the arctangent remains within a single branch and may be evaluated using the two argument atan2 function.

Eq. (12) fully supports multimode oscillation subject only to the phase being the sole state variable. If the effect on the dynamics of fluctuations of the magnitude is non-negligible recourse to the complex envelope model is necessary.

If sustained multimode oscillation is of interest, then the winner-takes-all competition may be suppressed by substituting the bell-shaped passband RF resonator by a broad flat-topped passband RF resonator or even by removing the RF resonator entirely. The impulse response $h$ may be then approximated by a Dirac distribution and (12) reduced to:

$$\theta_v(t) - \theta_v(t - \tau_D) = \phi + \left(\tan^{-1}\left(\frac{\rho \sin(\theta_w - \theta_v)}{1 + \rho \cos(\theta_w - \theta_v)}\right)\right)(t - \tau_D) \tag{14}$$

The free oscillator (static tuning, no injection or Langevin forcing terms) is then described by:

$$\theta_v(t + \tau_D) - \theta_v(t) = \phi \tag{15}$$

which admits solutions of the form:

$$\theta_v(t) = s(t) + \xi(t) \quad , \quad s(t + \tau_D) - s(t) = \phi \quad , \quad \xi(t + \tau_D) = \xi(t) \tag{16}$$

where $s$ is a staircase function with steps of height $\phi$ occurring at intervals of $\tau_D$ and $\xi$ is a periodic function with period $\tau_D$. Note that the staircase function may be expressed:

$$s(t) = \omega t + \tilde{\xi} \quad \omega \tau_D = \phi \tag{17}$$

where $\omega$ is the frequency shift common to all modes introduced by the tuning phase $\phi$ and $\tilde{\xi}$ is a zero mean sawtooth function with the same period as $\xi$. The periodic functions $\xi$ and $\tilde{\xi}$ are responsible for the appearance in the spectrum of a multitude of modes spaced in frequency by $1/\tau_D$.

If the initial contents of the delay line are set equal to $\xi$ over the interval $t \in [-\tau_D, 0)$, then a simulation of the free time delay oscillator with no RF resonator and $\phi = 0$ will reconstruct the periodic function $\xi$ for all $t \geq 0$. For $\phi \neq 0$ the staircase function adds to the solution by linearity. This generalizes so that simulations of (12) may be initiated in a constant magnitude but otherwise arbitrary multimode oscillation state by an appropriate choice of the initial contents of the delay line. In the sequel it is assumed unless otherwise stated that the oscillation has evolved to a predominately single-mode state prior to the onset of injection.

*C. Locked oscillation states*

Consider the injection of a pure carrier with frequency $\omega_i$ which gives rise to the phase ramp.

$$\theta_w(t) = \omega_i t \tag{18}$$

A phase locked solution of (14) corresponds to a phase ramp of the same slope but offset by a constant:

$$\theta_v(t) = -\theta_\infty + \omega_i t \tag{19}$$

Substituting (19) into (14) using $\omega_p \tau_D = \phi$ yields:

$$\tan\left((\omega_i - \omega_p)\tau_D\right) = \frac{\rho \sin(\theta_\infty)}{1 + \rho \cos(\theta_\infty)} \tag{20}$$

Consequently, there is a range of injection frequencies about *every* natural frequency:

$$(\omega_i - \omega_p)\tau_D \in [-\sin^{-1}(\rho), \sin^{-1}(\rho)] \tag{21}$$

for which a locked oscillation state exists.

### D. Serrodyne oscillation states

Suppose the frequency of the injected carrier $\omega_i$ falls within the locking range of a mode so that:

$$(\omega_i - \omega_p)\tau \in 2m\pi + [-\sin^{-1}(\rho), \sin^{-1}(\rho)] \quad ; \quad m \in \mathbb{Z} \tag{22}$$

The case $m = 0$ with $\rho < 1$ corresponds to injection within the locking range of the principal mode already considered. $m \neq 0$ corresponds to injection within the locking range of a mode other than the principal mode. The linear ramp solution corresponding to an injection-locked oscillation of the sidemode within locking range of the injected carrier frequency can be reached only if $\rho > 1$ and the arctangent is unwrapped. The latter is a reasonable expedient given that for $\rho > 1$ the trajectory of the vector sum of the oscillation and injection complex envelopes encircles the origin in the complex plane. For $\rho < 1$ the trajectory does not encircle the origin and no phase locked solution exists for $m \neq 0$.

Phase model simulations demonstrate that a solution of (14) given by:

$$\theta_v(t) = -\theta_\infty + \varpi_i t + \mathcal{S}(t) \tag{23}$$

is attractive, where $\mathcal{S}$ is a *sawtooth function* which has a peak-to-peak magnitude $2\pi$, a fundamental period $\tau$ and, where it exists, a derivative given by:

$$\tau \frac{d\mathcal{S}}{dt} = 2m\pi \quad ; \quad m \in \mathbb{Z} \tag{24}$$

It follows that $m$ cycles of the sawtooth function occur within each delay interval $\tau_D$. Now:

$$\theta_v(t) - \theta_v(t - \tau) = \varpi_i \tau \tag{25}$$

and, where it exists, the derivative:

$$\tau \frac{d}{dt}(\theta_w - \theta_v) = (\omega_i - \varpi_i)\tau - 2m\pi = 0 \tag{26}$$

which identifies $\varpi_i \tau$ as the remainder of $\omega_i \tau$ modulo $2\pi$. Where the derivative does not exist $\theta_w - \theta_v$ steps by $2\pi$ *instantaneously*. It follows that:

$$\theta_w - \theta_v = \theta_\infty \mod 2\pi \tag{27}$$

Consequently:

$$(\varpi_i - \omega_0)\tau = \tan^{-1}\left(\frac{\rho \sin(\theta_\infty)}{1 + \rho \cos(\theta_\infty)}\right) \quad ; \quad \omega_0 \tau = \phi \tag{28}$$

The serrodyne frequency-shifted oscillation is thereby phase locked to the injected carrier. In practice the fly-back of the sawtooth phase modulation cannot be instantaneous. The RF BPF ensures a short but continuous flyback transient occurs. The duration of the transient is inversely proportional to the RF BPF bandwidth. In simulations, a 3.6 MHz or 36 MHz bandwidth filter results in a transient of 2 μs or 200 ns respectively.

The serrodyne waveform consists of a concatenation of regular domains of a substantially pure oscillation phase locked to the injected carrier with the flyback transients forming the domain walls. The spectrum of such a waveform has a dominant component at the frequency of injection. It is observed experimentally that injection can be applied to select any mode among the multitude of natural modes [29]. The question then arises whether the serrodyne state can evolve into a locked pure oscillation state. The phase models that invoke the Leeson model show no evidence that the serrodyne state is other than persistent. However, the validity of the Leeson approximation is suspect for short transients. Envelope model simulations show that for a single-mode initial condition and for detuning within a subset of the locking range, the serrodyne waveform decays into a pure oscillation state.

### III. DESCRIPTION OF THE SIMULATION MODEL

Time domain simulations are performed using the Simulink™ block diagram environment. The phase noise spectrum is assessed by appropriate spectral analysis of the time domain data. It is the complex envelope of a pristine carrier corresponding to some nominal frequency of oscillation that is simulated. Since the nominal carrier is known completely, it conveys no information and is omitted. This is equivalent to translating the frequency origin to the nominal oscillation frequency. The complex envelope representation is motivated by the sample rate required in a digital simulation to avoid aliasing. For example, if the spectrum of the complex envelope of interest extends from -1 MHz to 1 MHz a sample rate of at least 2 MHz is required whereas to adequately represent the *same* signal with an explicit 10 GHz carrier requires a sampling rate of greater than 20 GHz. The reduced phase-only models are essentially equivalent to the envelope models but with the complex envelope magnitude held constant, thereby restricting the simulation to the evolution of the phase of an established oscillation. A phase model provides the option to invoke the Leeson approximation to the action on the phase of an RF BPF [27, 28]. The Leeson approximation has the merit of linearity, simplicity, and general utility, but its validity is suspect when spiking phenomena may occur.

The device of a complex envelope representation is used with advantage in optical circuit simulations [23] to avoid having to sample an optical carrier which has a frequency of the order of 200 THz. It is tempting to use an optical circuit simulator to simulate an OEO to take advantage of the sophisticated models for lasers, modulators, optical fibres, photodetectors, and a more limited range of electronic components. However, the number of samples required to adequately represent the RF modulation in transit through a fibre of length of the order of 10 km is prodigious (~1,000,000) and simulations of oscillators with coil lengths of up to ~ 100 m only is possible before exhausting available computing resources. This problem is avoided by treating the RF-photonic link as an RF delay line; indeed, the purpose of the photonics is to transport the RF carrier over a long path taking advantage of the low loss of the optical fibre. Ideally, the photonics could otherwise be ignored. However, the photonics contributes by a variety of mechanisms to RF phase & amplitude fluctuations which are included in the simulations. These observations reaffirm the rationale for using Simulink™ rather than an optical circuit simulator for the time domain simulations. Nevertheless, in contrast to the abstract description of the mathematical models, the Simulink™ models are built from blocks that correspond closely to the physical components that compose an OEO thereby providing intuition.

## A. Complex envelope simulation model

A top level (level 0) Simulink™ test harness for an envelope model of a single-loop OEO under injection by an external source is shown in Figure 2 (a). The simulation time is considered to have units of $1\ \mu s$. Figure 2(b) & (c) expand the *OEO* and the *RF source* to level 1 to reveal their respective internal structure. The constituent blocks are further expanded to reveal their contents in Figure 3 & Figure 4 at level 2 and Figure 3(f) & Figure 4(d) at level 3.

The OEO subsystem consists of an *Injection*, *Phase fluctuations*, *Initial condition*, *RF delay line*, *RF resonator*, *RF amplifier*, and a *Phase bias* subsystem arranged into a loop (Figure 3). These subsystems implement Eq. (5) with the addition of noise and an explicit initial condition.

The linear superposition $u = v + w$ of the injected carrier $w$ and the oscillation $v$ is implemented by the *Injection* subsystem (Figure 4 (a)) with provision to initiate injection with a specified injection ratio (*Gain* block) at a specified simulation time (*Switch* and *Step* blocks).

The delay operator $D_{\tau_D}$ is implemented by the *RF delay line* subsystem which makes use of the supplied *Transport Delay* block (Figure 3 (a)). The latter does not support complex scalar input and output variables, but it does support multidimensional real vectors. Consequently, the *Transport Delay* block is placed between a subsystem *Convert 1*, which maps the incoming complex envelope to a 2-dimensional real vector, and a subsystem *Convert 2*, which performs the corresponding inverse map.

The convolution operator $h \otimes (\cdot)$ is implemented by the *RF resonator* subsystem which makes use of the supplied *State Space* block (Figure 3 (b)). Since the latter does not support complex scalars but does support multidimensional real vectors, it is placed between *Convert 1* & *Convert 2* subsystems. A single pole LPF baseband equivalent to a BPF corresponds to the settings:

$$\boldsymbol{A} = -\boldsymbol{I}/\tau_R$$
$$\boldsymbol{B} = \boldsymbol{I}/\tau_R$$
$$\boldsymbol{C} = \boldsymbol{I}$$
$$\boldsymbol{D} = \boldsymbol{0}$$

where $\boldsymbol{I}$ is the $2 \times 2$ real identity, $\boldsymbol{0}$ is the $2 \times 2$ real zero matrix and $\tau_R$ is the on-resonance group delay of the RF resonator.

The nonlinear operator $K$ is implemented by the *RF amplifier* subsystem (Figure 3 (a)). The subsystem may be used to simulate an RF power amplifier with a memoryless limiting gain saturation mechanism. The complex envelope at the output is equal to the product of the input complex envelope and a real scalar saturable gain $\kappa$. The phase of the output is identical to the input, but the magnitude of the output is a monotonic increasing function of the magnitude of the input with an asymptote for small signals equal to the linear gain parameter and a least upper bound for large signals equal to the saturation parameter. The user-defined MATLAB Function code (Figure 3(f)) implements Eq. (11) to provide a saturable gain with unit linear gain and a unit magnitude maximum output. The *Product* and *Constant* blocks enable the linear gain parameter and saturation parameter of the RF amplifier to be set independently to arbitrary values.

The tuning phase $\phi$ is implemented by *Phase bias* subsystem (Figure 4(b)). The *Product* block provides a complex envelope at the output equal to the product of the input complex envelope and the complex phase factor $\exp(i\phi)$ supplied by the *Trigonometrical function* and *Constant* blocks.

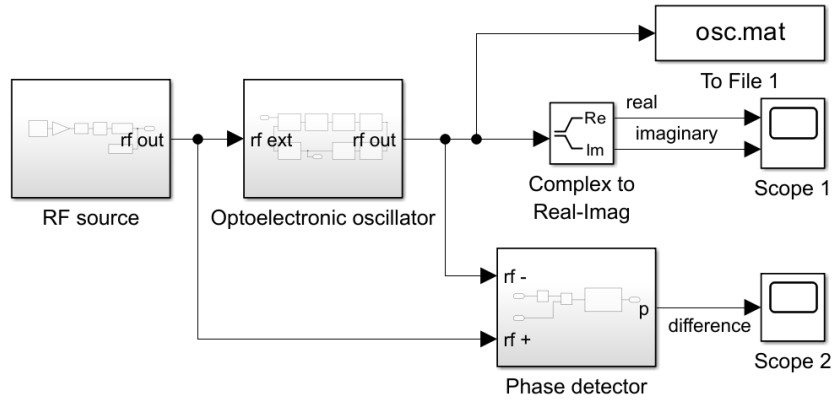

(a) *OEO* subsystem test harness

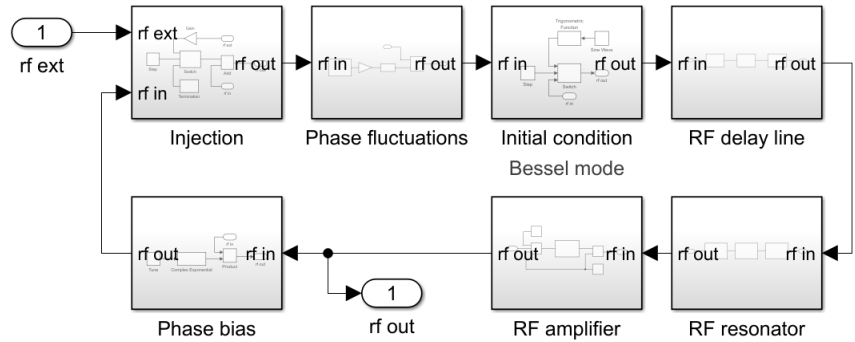

(b) *OEO* subsystem

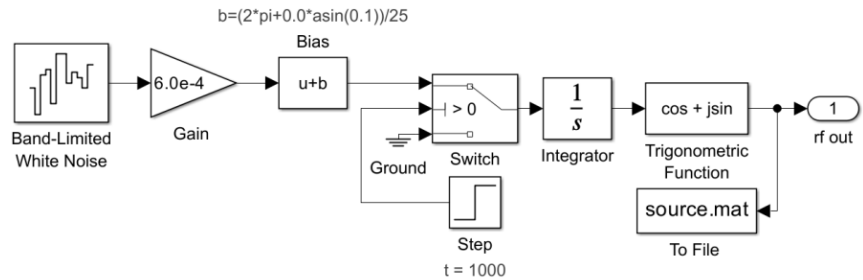

(c) *RF source* subsystem

Figure 2. Simulink™ envelope model of an optoelectronic oscillator under external injection by an RF source. (a) Test harness: Scope 1 measures the time evolution of the real and imaginary parts of the oscillation; Scope 2 measures the time evolution of the unwrapped phase difference between the output of the RF injection source and the oscillator. The To File block records the complex data for subsequent spectral analysis. (b) & (c) Level 1 expansion of the Optoelectronic oscillator and RF source subsystems.

The solution of the DDE describing a TDO requires knowledge of the entire initial contents of the loop. In this case, the set of all initial conditions forms an infinite dimensional vector space of functions over the interval $t \in [-\tau_G, 0)$ ; $\tau_G = \tau_D + \tau_R$ characteristic of a multimode oscillator with an infinity of modes. The purpose of the *Initial condition* subsystem is to load the loop with the requisite function.

Within the *Initial condition (Bessel mode)* subsystem (Figure 3(c)), the *Sine Wave* block together with the *Trigonometric Function* block generates the complex envelope of a carrier phase-modulated by a sinusoid defined by the *frequency*, *amplitude*, and *bias* parameters set within the *Sine Wave* block. At the start of simulation, the *Switch* block connects the complex envelope signal generated to the output port until the *Step* block changes state at the time set by its *Step Time* parameter when the *Switch* block connects the input port directly to the output port. The *Step Time parameter* should be set equal to the delay time. An alternative *Initial Condition (White noise)* subsystem (Figure 3(d)) loads the loop with a random complex envelope with Gaussian distributed real and imaginary parts.

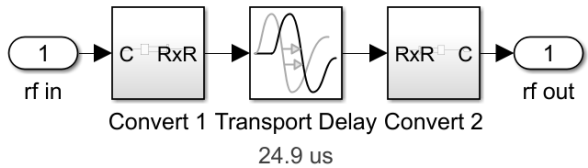
(a) *RF Delay* subsystem

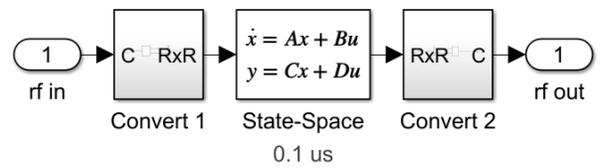
(b) *RF Resonator* subsystem

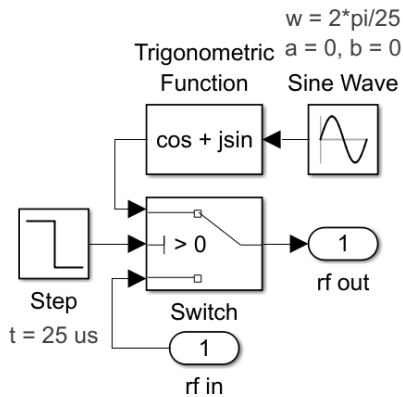
(c) *Initial condition (Bessel mode)* subsystem

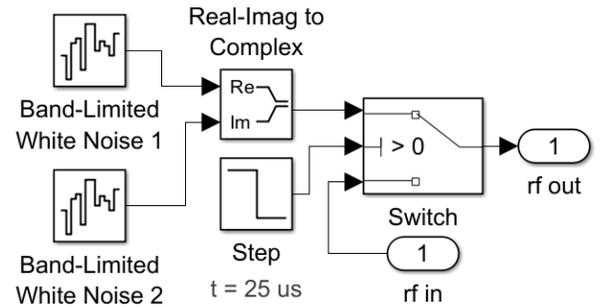
(d) *Initial condition (white noise)* subsystem

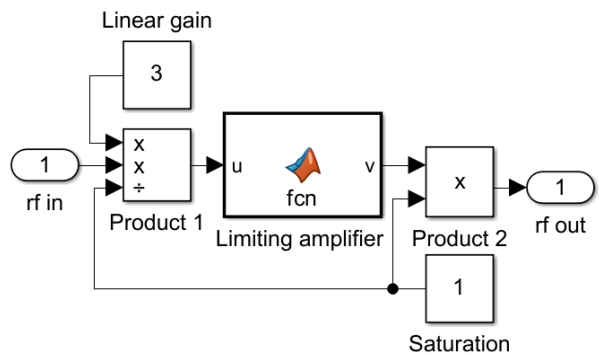
(e) *RF amplifier* subsystem

(f) *MATLAB fcn* block code

```
function v = fcn(u)
% Limiting amplifier
% gain saturation model
% unit linear gain
% unit saturation level

a = (4/pi)*abs(u);
if a<=1
    v = u;
else
    th = 2*acos(1/a);
    v = (1 - (th - sin(th))/pi)*u;
end
```

*Figure 3. Content of RF Delay, RF Resonator, Initial condition & RF Amplifier subsystems in Figure 2(b).*

The *Initial Condition* subsystems offer initial conditions corresponding to established single mode or highly multimode oscillation states. The *Initial Condition (Bessel mode)* subsystem is particularly useful when a frequency comb initial oscillation is desired and the *Initial Condition (White noise)* subsystem is particularly useful when either a realistic start-up from low noise is of interest or, at the other extreme, a highly random multimode initial condition is desired.

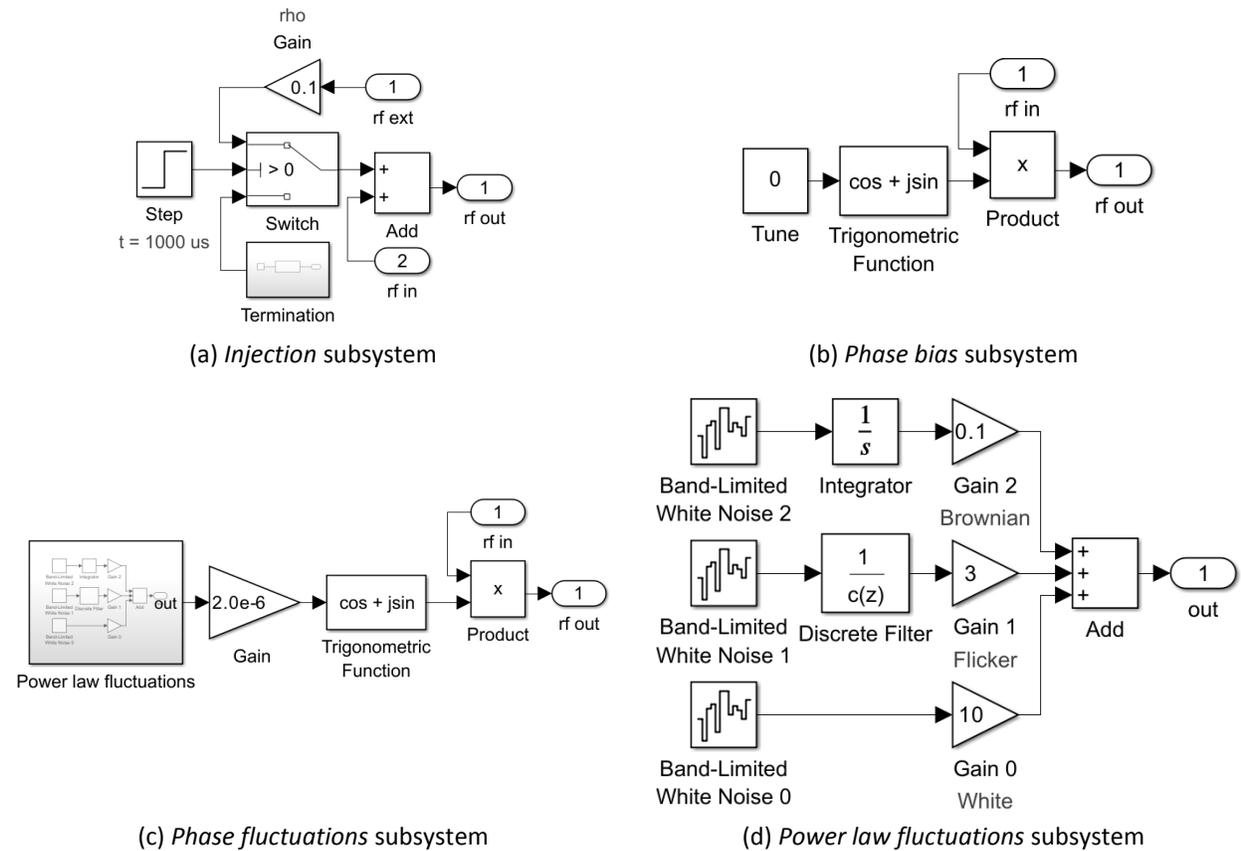

(a) *Injection* subsystem

(b) *Phase bias* subsystem

(c) *Phase fluctuations* subsystem

(d) *Power law fluctuations* subsystem

*Figure 4. Content of Injection, Phase bias, Phase fluctuations subsystems depicted in Figure 2(b)*

The *Phase fluctuations* subsystem (Figure 4(c)) perturbs the oscillation phase within the loop thereby generating oscillator phase noise. The *Product* block and *Trigonometric function* block form a phase modulator that is driven via the *Gain* block by a power law stochastic process generated by the *Power law fluctuations* subsystem. The *Gain* block permits adjustment of the overall level of the fluctuations while preserving the profile of the fluctuation spectral density.

Figure 4(d) illustrates the internal structure of the *Power law fluctuations* subsystem. There are three arms (0, 1, 2) that respectively generate white ($f^0$), flicker ($f^{-1}$), and Brownian fluctuations ($f^{-2}$) by filtering the independent identically distributed zero mean unit power Gaussian noise processes generated by the *Bandlimited White Noise* blocks. A weighted sum of spectrally shaped noise from each arm forms the output the *Power law fluctuations* subsystem. The weights provided by the *Gain* blocks are chosen to reproduce representative experimentally measured OEO phase noise spectral densities. The white noise arm (0) applies no filtering. The Brownian arm (2) applies a simple integrator to generate a Brownian motion stochastic process. The flicker noise arm (1) uses the *Discrete Filter* block as an infinite impulse response filter according to a method introduced by Kasdin [30]. To obtain a spectral density that follows $1/f$ power law over a frequency range spanning 3-4 decades, it is found that 30,000 filter coefficients are required. These are calculated by MATLAB code (Figure 5) stored in the model workspace. The code also generates and stores in the model workspace an array of random seeds for the *Band-Limited White Noise* blocks ensuring independence.

```matlab
% discrete filter coefficients
m = 30000;
c = zeros(1,m);
c(1) = 1;
for k=1:(m-1)
    c(k+1) = ((2*k-3)/(2*k))*c(k);
end

% white noise block seeds
rng('default'); % reset global stream
s = randi(10000,[1,10]);
```

*Figure 5 Model workspace MATLAB code to initialise a $1 \times m$ array of discrete filter coefficients and a $1 \times N$ array of random seeds. N should at least equal the number of Bandlimited White Noise blocks used in the model. In this example $m = 30000$ and $N = 5$.*

The injected carrier $w$ is supplied by the *RF source* subsystem (Figure 2(a)) which is modelled as a simple phase integrator. The complex envelope is formed by complex exponentiation by a *Trigonometric Function* block of the output of an *Integrator* block driven by a *Bandlimited White Noise Block* weighted by a *Gain* block with an additive constant provided by a *Bias* block (Figure 2(c)). Provision is made using the *Switch* block to defer the activation of the RF source to the *step time* parameter of the *Step* block. The weight provided by the *Gain* block is chosen to reproduce representative measured phase noise spectral densities of dielectric resonator oscillator (DRO) based stable local oscillators (STALO). For zero oscillator tuning phase ($\phi = 0$), the *Bias* block parameter $b$ corresponds to the detuning $\omega$ of the injection source from the natural frequency of the oscillator. For $\phi = 0$, $\rho = 0.1$, $\tau_G = 25\ \mu s$, setting $b = 0.9 \sin^{-1}(0.1)/25$ corresponds to injection detuned by 90% of the high frequency limit of the locking range defined by Eq. (21) of the principal ($m = 0$) mode and $b = 2\pi$ corresponds to injection with zero detuning from the adjacent sidemode ($m = 1$).

## B. Reduced phase-only simulation model

A top level (level 0) Simulink™ test harness for a reduced phase-only model of a single-loop OEO under injection by an external source is shown in Figure 6(a). The *RF source* and the *OEO subsystems* are expanded to reveal their respective internal (level 1) structure in Figure 6 (b) & (c). The constituent blocks are further expanded to reveal their (level 2) contents in Figure 6 (d), (e) & (f). The *Power law fluctuations* subsystem in Figure 6 (e) is identical to that used in the complex envelope models (see Figure 4(c) & (d)).

The *OEO* subsystem consists of an *Injection* subsystem & *Bias (Tune)* block, *Phase fluctuations* & *Initial condition* subsystems, and *Transport Delay* & *Transfer Fnc (RF resonator)* blocks arranged into a loop (Figure 6(c)). These subsystems and blocks implement Eq. (12) with the addition of noise and an explicit initial condition. The injection induced phase shift of the oscillation (i.e., the arctangent term in Eq. (12)) is provided by the *Injection* subsystem (Figure 6(d)) with provision to initiate injection with a specified injection ratio at a specified simulation time (*Switch* and *Step* blocks). In this example, the step occurs at $t = 1000\ \mu s$ and is of height $\rho = 0.1$. The delay operator $D_{\tau_D}$ is implemented by the *Transport Delay* block; the convolution operator $h \otimes (\cdot)$ is implemented by the *Transfer Fcn* (*RF resonator*) block; and the tuning phase $\phi$ is implemented by *Bias (Tune)* block.

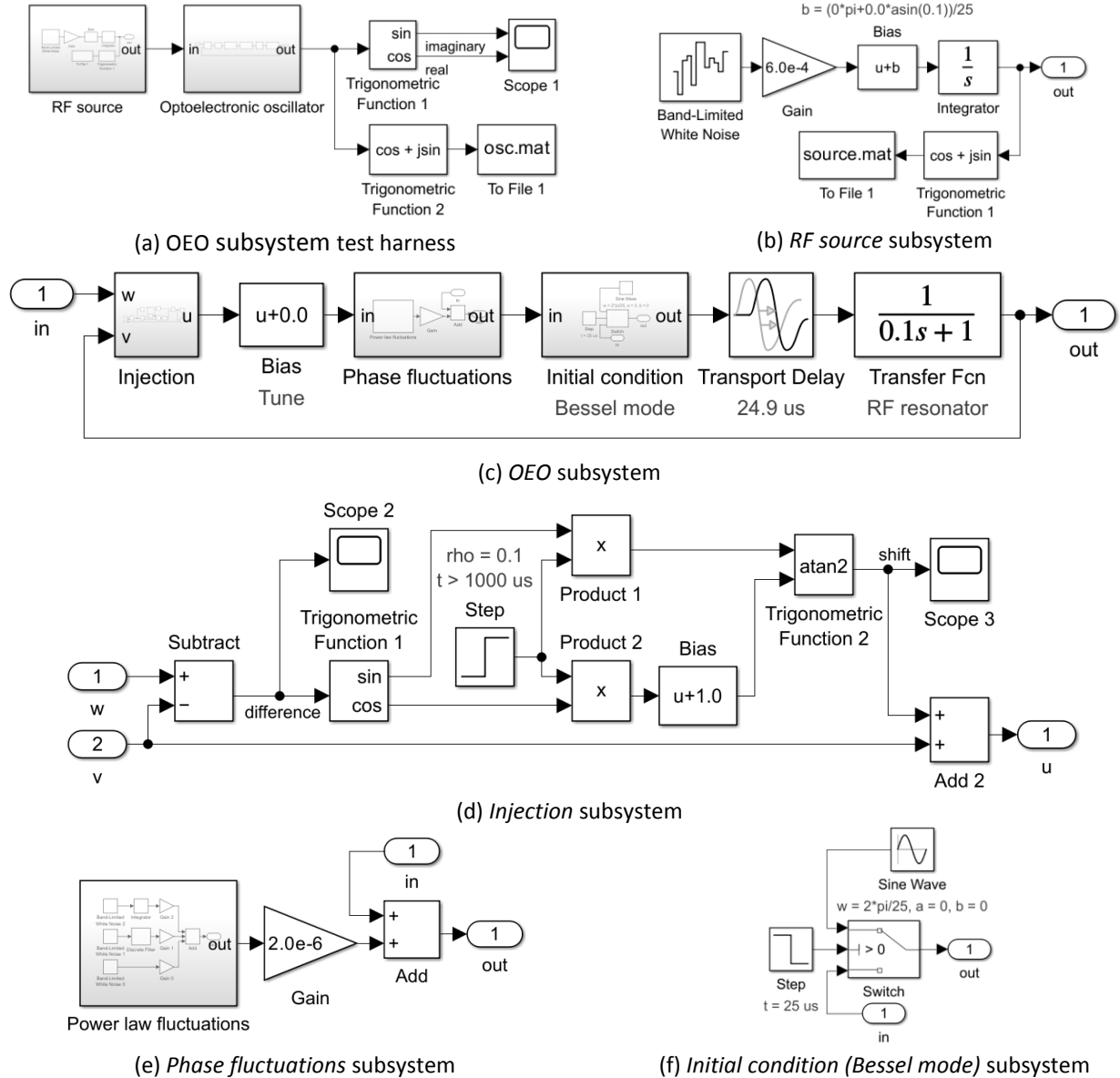

(a) OEO subsystem test harness

(b) *RF source* subsystem

(c) *OEO* subsystem

(d) *Injection* subsystem

(e) *Phase fluctuations* subsystem

(f) *Initial condition (Bessel mode)* subsystem

*Figure 6. Simulink™ phase model of an OEO under external injection by an RF source. (a) Test harness: Scope 1 measures the time evolution of the real and imaginary parts of the oscillation; the To File blocks record the complex envelope associated with the oscillator phase for subsequent spectral analysis. (b)&(c) Level 1 expansion of the RF source and Optoelectronic oscillator subsystems. (d), (e) & (f) Level 2 expansions of the Injection, Phase fluctuations & Initial condition (Bessel mode) subsystems. The Power law fluctuations subsystem in (e) is expanded in Figure 4(d).*

The *Injection* subsystem is the only source of nonlinearity when the Leeson approximation is invoked. For small phase fluctuations it is found that, in a large neighbourhood of $\theta_w - \theta_v = 0$, the injection subsystem is well approximated by the linear superposition:

$$\theta_u = \eta \theta_w + (1 - \eta)\theta_v \quad ; \quad \eta = \frac{\rho}{1 + \rho}$$

i.e., the *Injection* subsystem reduces to a simple linear coupler. The linearized *OEO* subsystem is amenable to tractable analysis. In particular, the phase noise spectrum may be derived analytically.

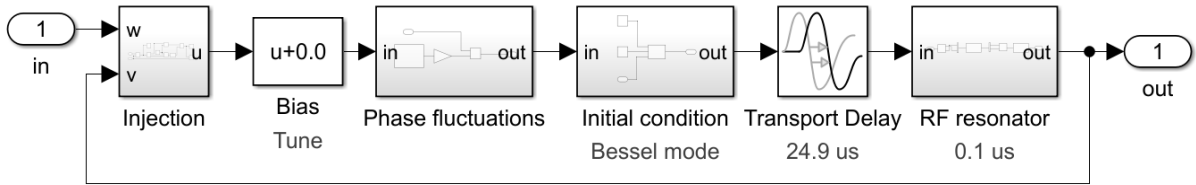

(a) *Optoelectronic oscillator* subsystem with *Transfer Fnc* block replaced by custom *RF resonator* subsystem.

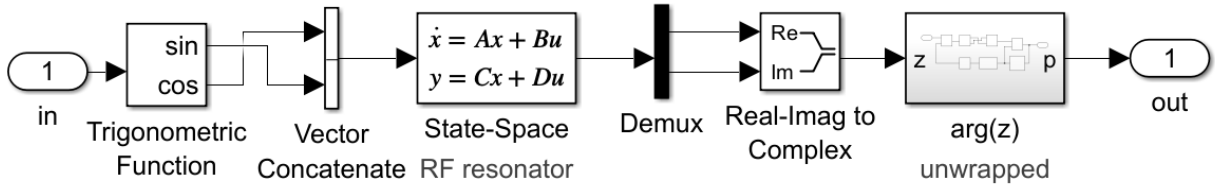

(b) Custom *RF resonator* subsystem

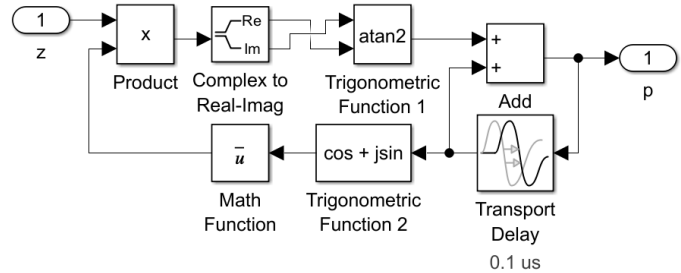

(c) *arg(z) (unwrapped)* subsystem

*Figure 7 Alternative Simulink™ phase model not invoking the Leeson approximation. (a) Level 1 contents of the Optoelectronic oscillator subsystem with a custom RF resonator subsystem replacing the Transfer Fnc block in Figure 6(c). (b) Level 2 contents of the custom RF resonator subsystem. (c) Level 3 contents of the arg(z) (unwrapped) subsystem.*

If the validity of the Leeson approximation is suspect, the *Transfer Fnc (RF resonator)* subsystem in Figure 6(c) may be replaced by the custom *RF resonator* subsystem shown at level 1 in Figure 7 (a) and expanded to level 2 & 3 in Figure 7(a) & (b). A *State-Space* implementation of a complex envelope model of the RF resonator is placed between a *Trigonometric Function* & *Vector Concatenate* block that generate a 2-dimensional real vector representation of a unit magnitude complex envelope and a *Demux* & *Real-Imag to Complex* block that reconstruct the complex envelope *z* as filtered by the resonator. The *arg(z) (unwrapped)* subsystem provides a continuous phase output provided the magnitude of the complex envelope never passes through zero. The unwrapping is achieved by using a short time delay to provide a previous value of the unwrapped phase which is used to avoid evaluating the arctangent near its branch cut (Figure 7(c)). The custom *RF resonator* subsystem is useful in distinguishing between phenomena due to failure of the Leeson model in reduced phase models from those due to loss of amplifier saturation in complex envelope models.

IV. SIMULATION RESULTS

Representative results generated by the complex envelope and reduced phase-only Simulink™ simulation models are presented. The results are selected to illustrate that the DDE formalism of injection locking captures OEO multimode oscillation phenomena. Table 1 lists the principal simulation parameters and their values used unless otherwise stated in the text.

| | |
|---|---|
| $\tau_D = 24.9\ \mu s$ | Delay line delay time |
| $\tau_R = 0.1\ \mu s$ | Resonator on-resonance group delay |
| $\tau_G = \tau_D + \tau_R = 25\ \mu s$ | Round trip group delay time |
| $\Delta f = 1/\pi\tau_R = 3.18\ MHz$ | Resonator bandwidth |
| $FSR = 1/\tau_G = 40\ kHz$ | Frequency interval between modes |
| $\rho = 0.1$ | Injection ratio |
| $m = 0$ | Mode index |
| $\omega\tau_G = 2m\pi + c\sin^{-1}(\rho)$<br>$c \in [-1,1]$ | Detuning within locking range of mode $m$ |

*Table 1 Principal simulation model parameters*

An OEO with a wide flat-topped passband RF BPF or no RF BPF supports concurrent oscillation of a multitude of modes. Figure 8(a) shows results of a complex envelope simulation of a free running OEO using the model shown in Figure 2 with the *RF resonator* and *phase fluctuations* subsystems commented through and the *RF source* subsystem commented out. An established oscillation consisting of a multitude of modes of comparable magnitudes is observed in the spectrogram that persists indefinitely following start-up from noise provided by the *Initial condition (white noise)* subsystem (Figure 3(d)).

The bell-shaped transfer function of the RF resonator when in place results in weak damping of the sidemodes ultimately leading to a single-mode oscillating state via a winner-takes-all gain competition process (Figure 8). In the case of the white noise initial condition, the winning mode is close to but not necessarily at the passband centre due to the random nature of the initial condition. In Figure 8 (b), the first adjacent mode to the mode at zero offset frequency is the winner. In practice, the internal fluctuations of the oscillator will maintain a residual level of sidemodes.

Injection locking theory typically assumes an established single-mode initial oscillation. For comparison between theory and simulation, it is consequently convenient either to initiate injection after the free-running oscillator has reached a substantially single-mode oscillating state or to set a single-mode initial condition. In the case of an OEO with large delay it is customary to adapt the theory of Adler [31], Paciorek [32], Stover [33] & Armand [34] to the problem. The classical theory describes a *single mode* oscillator with a resonator parameterised by its quality factor $Q$ and natural frequency $\omega_0$. it is asserted (e.g., Fan et al [35]) that a delay line with delay time $\tau_D$ operating at a frequency $\omega_0$ has a quality factor $Q = \omega_0\tau_D/2$. The assignment of a $Q$ to a delay line is fraught [36] in terms of accepted notions [37]. In essence, the delay line and RF-resonator are discarded and replaced by an ultra-high $Q$ resonator with the same on-resonance delay.

Figure 9 shows a comparison of the RF and phase noise spectra of the oscillation generated by an injection locked OEO and, by this customary reasoning its associated classical analogue. The classical analogue seriously underestimates the phase noise beyond midway between the oscillating mode and its adjacent sidemode and completely fails to describe the sidemode structure. In this example, a complex envelope model is used but the phase model provides the same results. When linearized the phase model provides an analytic prediction of the phase noise spectrum that accurately fits the phase noise spectrum of the simulation data.

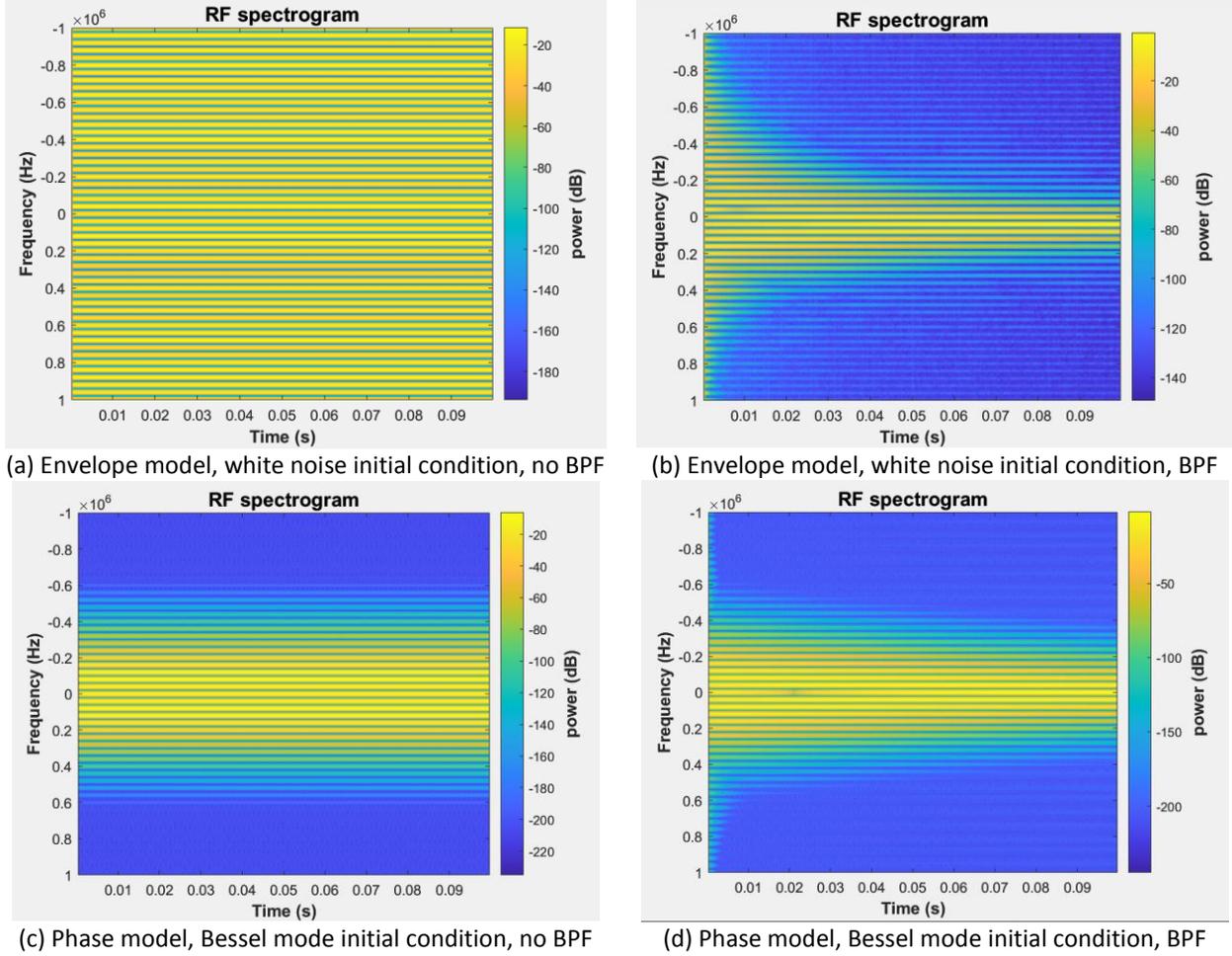

(a) Envelope model, white noise initial condition, no BPF
(b) Envelope model, white noise initial condition, BPF
(c) Phase model, Bessel mode initial condition, no BPF
(d) Phase model, Bessel mode initial condition, BPF

*Figure 8. A spectrogram of the oscillation of a free-running OEO with 25 μs delay generated by an envelope (upper panels) or phase model (lower panels) with (right panels) or without (left panels) the RF BPF. The resonant peaks are spaced by 40 kHz. The spectrogram colour map encodes the squared magnitude in dB. The oscillator is free of perturbations.*

Figure 10 shows the results of a simulation of the dynamical behaviour of an OEO under injection locking over a $5000 \, \mu s$ time interval. The OEO is prepared in a single-mode initial oscillation state. At $t = 1000 \, \mu s$, injection by the RF source is initiated. The injected carrier is detuned by 90% of the locking range ($\omega \tau_G = 0.9 \sin^{-1}(0.1)/25$). On this timescale, the response of the OEO oscillation to the onset of injection is substantially immediate (Figure 10(a)). The phase difference between the OEO and source evolves smoothly and monotonically to a steady phase-locked state (Figure 10(b)). The locking behavior is similar to a phase locked loop with proportional control [30]. In this example, an envelope model is used but a phase model provides identical results.

Figure 11 provides simulation results provided by the phase model (Leeson approximation) illustrating the serrodyne oscillation state that occurs for injection within the locking range of a sidemode to the oscillating mode. In the example, the injected carrier is tuned to the first adjacent sidemode ($\omega \tau_G = 2\pi$). The oscillation phase, shown in Figure 11(a), executes a sawtooth waveform with almost vertical $2\pi$ flyback transitions (cycle-slips) and a period $\tau_G = 25 \, \mu s$. If the injected carrier is tuned to sidemode $m$ the period of the sawtooth decreases to $\tau_G/m$, i.e., there are $m$ regularly spaced cycle-slips within the fundamental period $\tau_G$. The relative phase between the OEO and RF source, shown in Figure 11(a), executes a staircase waveform with steps of height $2\pi$. The almost horizontal steps demonstrate a phase-locked state between cycle-slips. The associated complex envelope, shown in Figure 11(a), features domains delimited by cycle-slip transitions containing a carrier at the same frequency as the RF source that extrapolates perfectly across the domain walls.

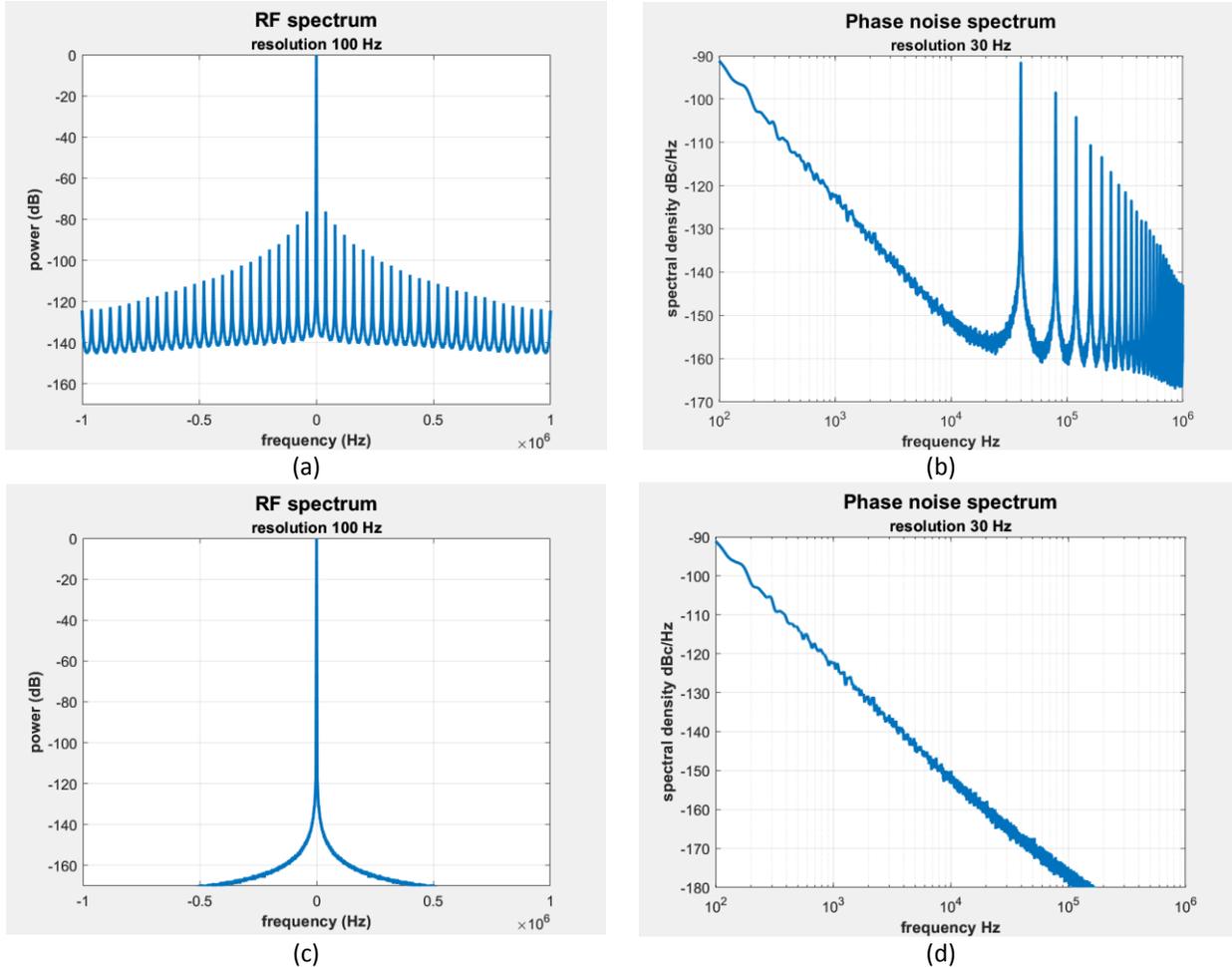

Figure 9. Comparison of the RF spectrum (left panels) and phase noise spectrum (right panels) provided by the complex envelope simulation model of an injection locked optoelectronic oscillator (upper panels) and an injection locked classical oscillator with the same on-resonance group delay (lower panels). The injection ratio and the phase fluctuation parameters within the source and oscillator are identical in the two cases. The customary adaption of classical injection locking theory to a time delay oscillator seriously underestimates the phase noise beyond midway between the locked mode and its adjacent sidemode and completely fails to describe the sidemode structure.

The detuning from the natural frequency of the sidemode adds a ramp of slope $\varpi_i$ to the sawtooth waveform so that the leading-edge slope remains $\omega_i$, and the appearance of the staircase waveform is unchanged. There are $m$ cycle-slips (domain walls) in each base period of $\tau_G$. Since $\omega\tau_G/2\pi$ is no longer an integer, the carrier translates slowly with respect to the domain walls. Phase model simulations show no evidence that the serrodyne state is other than persistent.

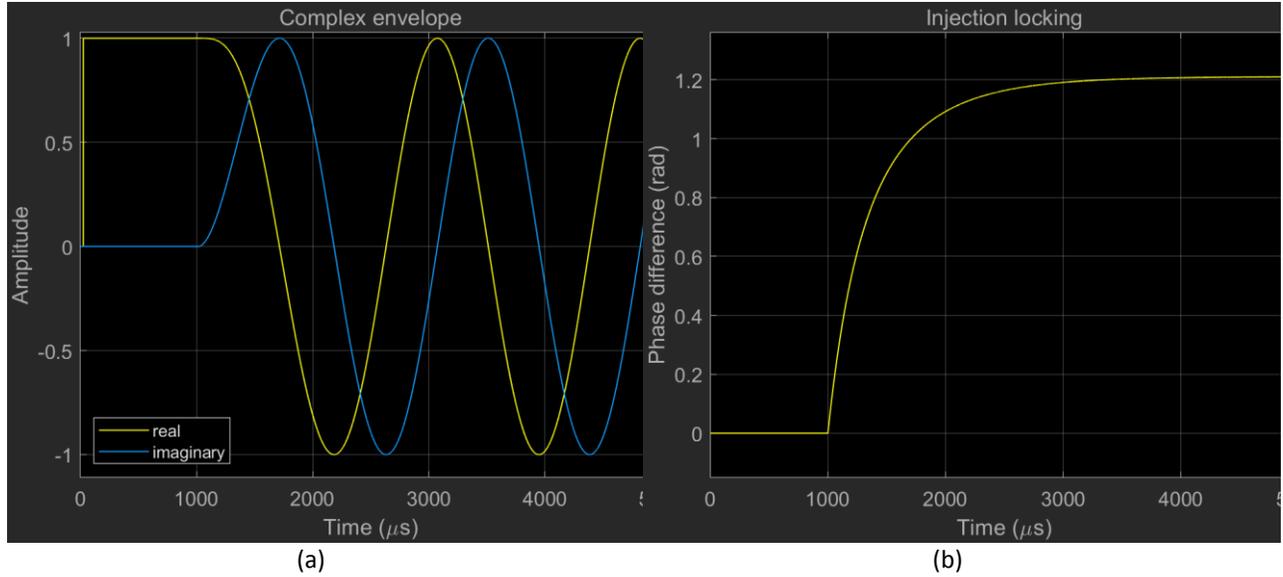

*Figure 10. (a) Evolution of the complex envelope of the oscillation of an OEO under injection. (b) The associated evolution of the phase difference between the OEO and the RF source. The OEO is prepared in a single mode initial oscillation state. The injection is initiated at $t = 1000\ \mu s$. The detuning is 90% of the locking range of the principal mode. The OEO dynamical behaviour is similar to a phase-locked loop with proportional control [30]. In this example an envelope simulation model is used but a phase model provides an identical result.*

Figure 12 shows envelope model simulation results featuring spiking behaviour. It is found in that the serrodyne oscillation may decay into a pure carrier with the same frequency as the injection, as shown in Figure 12(a), (b), (c). It persists only for detuning within a small interval within the locking range extending from the nearest edge to the principal mode. For example, when tuned to the first adjacent sidemode, it persists indefinitely only for $\omega\tau_G \in [2\pi - \sin^{-1}(0.1), 2\pi - 0.62\sin^{-1}(0.1)]$.

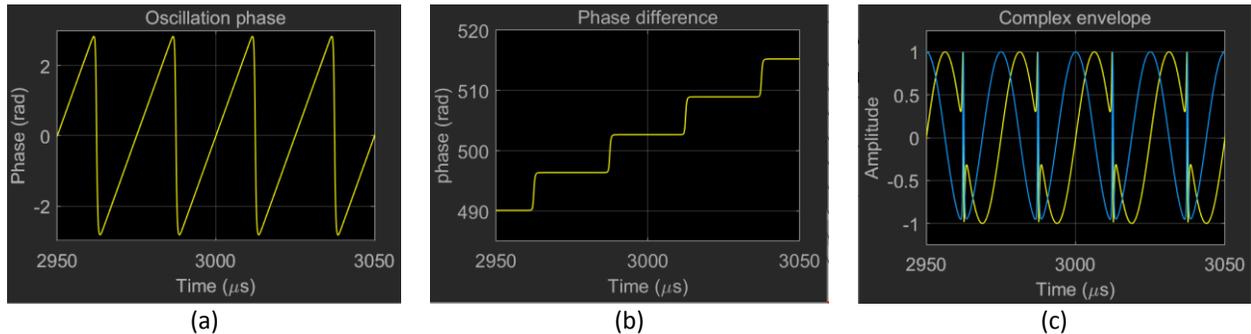

*Figure 11. Serrodyne oscillation state waveforms provided by the reduced phase-only simulation model. The injected carrier is tuned to the first adjacent sidemode ($\omega\tau_G = 2\pi/25$) (a) The oscillation phase executes a sawtooth waveform with an almost vertical $2\pi$ flyback (cycle-slip). (b) The oscillation phase relative to the source executes a staircase waveform with steps of height $2\pi$. The almost horizontal steps demonstrate a phase-locked state between the cycle-slips. (c) The associated complex envelope has domains containing a carrier at the frequency of the adjacent side mode and delimited by cycle-slip transitions. The waveform extrapolates perfectly across the domain walls.*

The transition from the serrodyne state to the pure oscillation state is heralded by the appearance of downward spikes in the magnitude of the RF resonator output, which can cause loss of amplifier saturation and hasten the decay. However, substitution of the exact phase model in place of the Leeson model of the RF resonator confirms that the fundamental cause of the decay is the failure of the Leeson approximation and not loss of amplifier saturation, as shown in Figure 12(d), (e), (f). The Leeson model of the RF filter acts on the phase (Figure 11 (a)), smoothing the $2\pi$ phase transients and limiting their minimum duration.

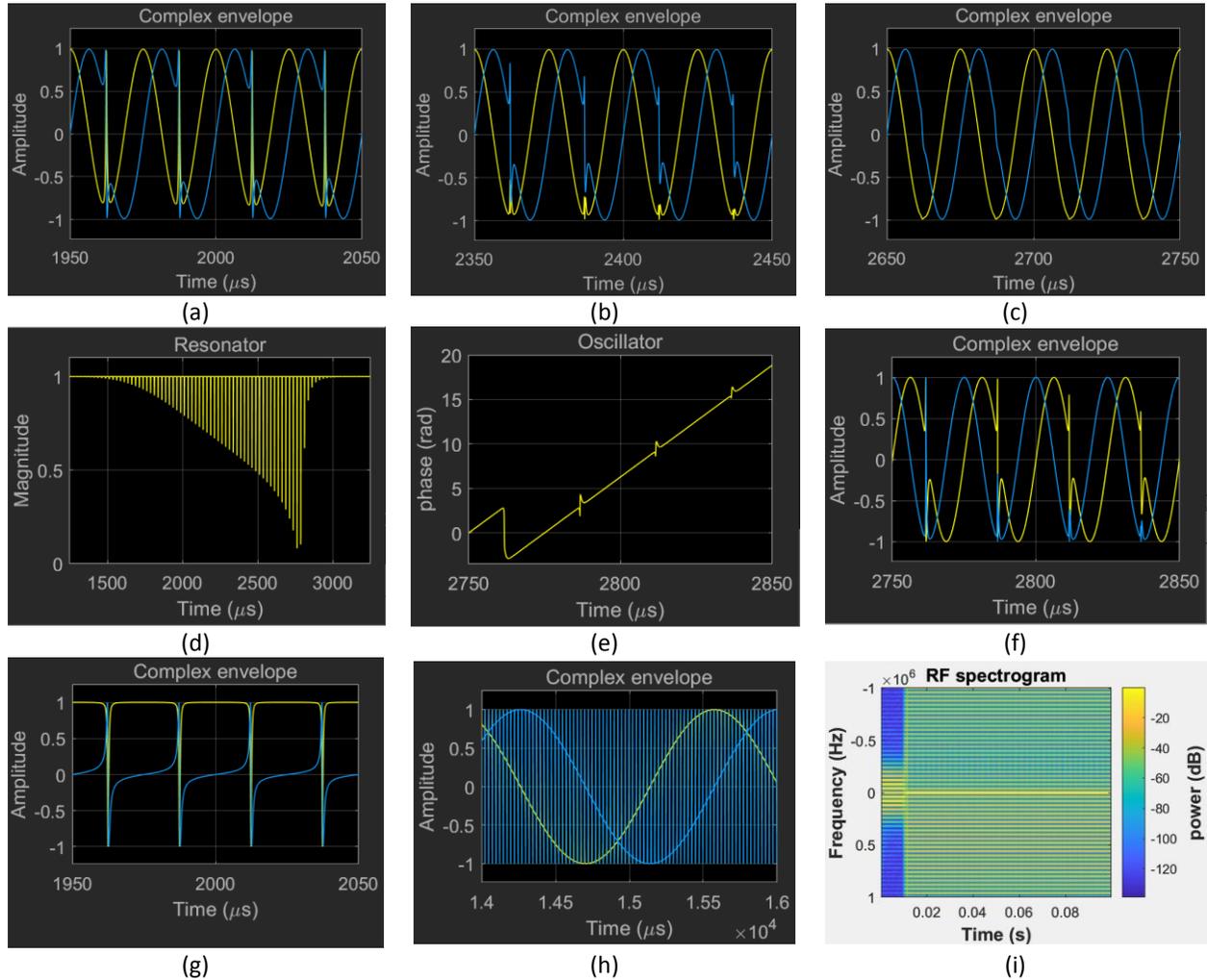

Figure 12 Serrodyne oscillation state waveforms provided by the complex envelope simulation model (a)-((g), (h) & (i) and by the reduced phase simulation model with exact RF resonator subsystem (d)-(f). In (a)-(g) the injected carrier is tuned to the first adjacent sidemode ($\omega\tau_G = 2\pi/25$). The serrodyne oscillation state can persists indefinitely for detuning within a small interval within the locking range extending from the nearest edge to the principal mode. Otherwise, it is observed to decay into a pure locked oscillation state as shown in (a)-((c). The decay is heralded by notches in the magnitude of the complex envelope which can cause loss of saturation but the exact phase model results (d)-(f) confirm that the fundamental cause is failure of the Leeson approximation. (g) illustrates that appearance of a plot of the real & imaginary parts of a complex envelope can be deceptive. Plot (a) and (g) illustrate the same oscillation state and are entirely equivalent. The difference is the nominal carrier frequency in (a) is set to the natural frequency of the main oscillation mode but in (g) it is set to the frequency of the injected carrier. (g) illustrates that the locked carrier is phase modulated by a regular train of short $2\pi$ phase pulses. In (h) & (i) the injected carrier is tuned to the edge of the locking range of the principal mode $\omega\tau_G = 0.9\ sin^{-1}(0.1)$. The multimode initial condition as used to generate Figure 8(c) & (d) is found to generate a$2\pi$ phase pulse train (h) that persist indefinitely as shown by the spectrogram (i).

The exact model of the RF filter acts on the complex envelope generated by the phase. For parameter regimes leading to decay, the flyback phase transient becomes steeper as the simulation proceeds; the relative proportion of time occupied by the transient to the locked carrier decreases, while the spectral content of the transient increasingly falls outside the passband of the RF filter. Consequently, the RF filter succeeds in erasing the narrow domain walls (Figure 12 (a), (b), (c), & Figure 12 (e), (f)) and the domains coalesce leaving only the locked carrier. This is consistent with the experimental observation that injection can be applied to select any mode among the multitude of natural modes [29].

In simulations, the multimode initial oscillation state used to generate Figure 8(c) & (d) also triggers the generation of a $2\pi$ phase pulse train by an OEO under injection tuned to the edge of the locking range of the nominal ($m = 0$) principal mode, as shown in Figure 12 (h) & (i). The spike train (Figure 12 (h)) generated by the complex envelope simulation model persist indefinitely while the phase model (Leeson approximation) generates a double spike train those decays. The replacement of the Leeson model by the exact model of the RF filter resolves the discrepancy.

The appearance of sinusoidal waveforms in between transients in all the plots of the real & imaginary parts of the complex envelope in Figure 11, Figure 12 and especially Figure 12 (h) may be misleading. The complex envelopes are defined relative to a nominal carrier which in the Simulink™ models presented is set to zero offset frequency. However, for locked serrodyne modes the carrier frequency is set by the RF source. Hence the pertinent complex envelope is $\exp(i\theta)$ where $\theta = \theta_v - \theta_w$ is the relative phase between oscillator and source which has a staircase waveform with $2\pi$ steps. The staircase waveform is transformed by the complex exponentiation into pure spike trains as shown in Figure 12(a). This figure was generated for the special case of zero detuning that results in an asymptotic phase shift $\theta_\infty = 0$. However, if $\theta_\infty$ is absorbed by a phase shift of the carrier then Figure 11 (c) and Figure 12(a), (b), (c), (f), & (h) would all have the same appearance as Figure 12(g), i.e.. a train of pure $2\pi$ pulses phase modulates a carrier locked to the injection source. The simulation results are consequently redolent of the recent experimental observations of $2\pi$ phase pulse trains reported by Diakonov et al [14].

A periodic train of narrow pulses necessarily requires the *coherent* superposition of a multitude of modes. In this sense the spiking phenomena may be considered a form of mode-locking. TDOs such as lasers and OEOs support a multitude of modes and consequently, spiking phenomena is to be expected. There is some literature on baseband spiking phenomena in broadband OEOs triggered by external pulses [15,16]. Mode-locked laser theory and practice is well advanced [17] and OEOs support analogous phenomena [18-22]. However, beyond the cycle-slips observed in injection pulling of classical oscillators [31,34,35], the theory and simulation of injected carrier induced spiking phenomena in a TDO appears not to have been explored.

V. SUMMARY & CONCLUSIONS.

A comprehensive Simulink™ simulation model of the dynamical behavior of an OEO under injection by an external source has been described. The model builds on the foundations of a previously reported delay DDE formulation of injection locking of TDO [4]. The model has two varieties, a complex envelope model which is relates more closely to the physics and a reduced phase only model that relates more closely to analytic solutions and classical theories. The correspondence between the blocks and the oscillator components offers intuition and considerable freedom to explore different circuit architectures and design variations with minimal coding effort. Selected simulation results demonstrate that the simulation models fully support multimode oscillation and correctly predict the phase noise spectral density. Moreover, new theoretical and simulation results have been presented of injection induced persistent spiking phenomena redolent of recent experimental observations of $2\pi$ phase pulse trains in a broadband OEO under injection [14].